\documentclass[aps,showpacs,prl,twocolumn]{revtex4}
\usepackage{amssymb,amsfonts,amsmath,epsfig,graphicx}

\newcommand{\be}{\begin{eqnarray}}
\newcommand{\ee}{\end{eqnarray}}
\newcommand{\bc}{\begin{center}}
\newcommand{\ec}{\end{center}}
\newcommand{\bea}{\begin{eqnarray}}
\newcommand{\eea}{\end{eqnarray}}

\newcommand{\beq}{\begin{equation}}
\newcommand{\eeq}{\end{equation}}

\newcommand{\nn}{\nonumber \\ }

\def\fun#1#2{\lower3.6pt\vbox{\baselineskip0pt\lineskip.9pt
\ialign{$\mathsurround=0pt#1\hfil##\hfil$\crcr#2\crcr\sim\crcr}}}

\begin{document}

\title{Some indication for a missing chiral partner $\eta_4$ around 2 GeV}

\author{L. ~Ya. ~Glozman$\,^{1}$ and A.~Sarantsev$\,^{2,3}$}
\affiliation{$^1\,$ Institute for Physics, Theoretical Physics Branch,
University of Graz, Universit\"atsplatz 5, A-8010 Graz, Austria
$^2\,$Helmholtz-Institut f\"ur Strahlen- und
Kernphysik,
Universit\"at Bonn, Germany\\
$^3\,$Petersburg Nuclear Physics Institute, Gatchina, Russia}

\begin{abstract}
The high-lying mesons in the light quark sector previously obtained
from the partial wave analysis of the proton-antiproton annihilation
in flight at 1.9 - 2.4 GeV region at CERN reveal a very high degree
of degeneracy. This degeneracy can be explained as due to an
effective restoration of both $SU(2)_L \times SU(2)_R$ and $U(1)_A$
symmetries combined with a principal quantum number $\sim n + J$. In
this case there must be chiral partners for the highest spin states
in the 2 and 2.3 GeV bands presently missing in the data. Here we
reanalyze the Crystal Barrel data and show an indication for
existence of the missing $4^{-+}$ state around 2 GeV. This result
calls for further experimental search of the missing states both in
the proton-antiproton annihilation and in the production reactions.
\end{abstract}

\maketitle

\section{Introduction}

Until 10 years ago  a little was known about  mesons in the light
quark sector with masses in the region of 2 GeV. A  development in
the field was promoted by a publication of 4 papers
\cite{A1,A2,A3,A4} that contained results of a partial wave analysis
of the proton-antiproton annihilation into mesons at LEAR (CERN) in
the energy range 1.9 - 2.4 GeV with four different sets of quantum
numbers. A lot of new mesons have been discovered. This is not
accidental because the proton-antiproton annihilation into mesons is
a formation experiment and consequently it allows for a systematic
exploration of the whole kinematical region. This is in contrast
with the production experiments where one typically looks for a
meson resonance in correlations of some secondary particles in
high-energy reactions. In the latter case it is difficult to explore
systematically a large kinematical region and a search was typically
guided by  predictions of theoretical models, such as the linear
Regge trajectories  model \cite{Regge}, or by the Goddfrey-Isgur
constituent quark model for mesons \cite{GI}. There is no other
systematic experiment in the same kinematical region so all these
new resonances await for their confirmation before they penetrate
into a Meson Summary Table of Particle Data Group (they are listed
as "Other Light Mesons") \cite{PDG}. Nevertheless some of these new
mesons are regarded by the authors as very reliable, because they
are seen at least in a few independent decay channels in data with
polarization, while the other are less reliable \cite{BUGG}. A
striking feature of these data is that they reveal a high degree of
degeneracy, namely states with different spins, parities and
isospins "perfectly group into two clusters around the masses of
$\simeq 2$ GeV and of $\simeq 2.2 - 2.3$ GeV" \cite{G1}, see also
figures in refs.\cite{AF1,G2,KZ} and Fig. 1 below. Such a degeneracy
indicates a symmetry. Understanding a source of this symmetry would
clarify a fundamental question of mass generation in QCD,
 an interconnection of confinement and chiral symmetry breaking and physics responsible
 for the angular momentum generation.

\begin{figure}[pt]
\includegraphics[width=0.48\textwidth,clip=on]
{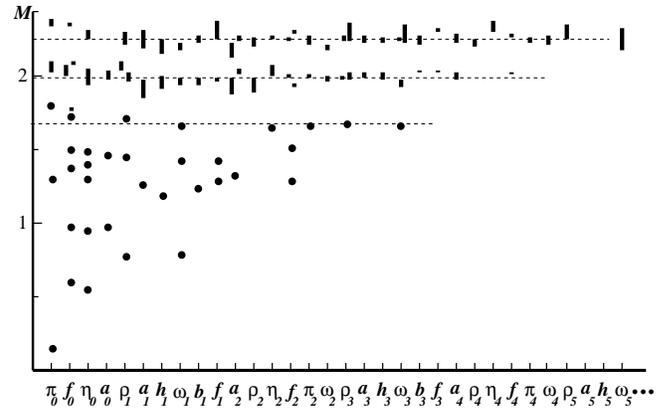}
\caption{\label{spectrum} Masses (in GeV) of the well established
states from PDG (circles) and new $\bar n n$ states from the
proton-antiproton annihilation at LEAR (strips)}
\end{figure}

 There are two different scenarios. The first one is based on the conjecture of
 effective chiral restoration in highly excited hadrons \cite{G3,CG,G2,C} which
 was promoted earlier given  data for established highly excited baryons. The new high-lying
 mesons have been analyzed in  refs. \cite{G4,G5} and the analysis has revealed that the
 data is well consistent with the conjecture of effective restoration of both
 $SU(2)_L \times SU(2)_R$ and $U(1)_A$ symmetries. It is well seen from Fig. 1, however,
 that the chiral partners are missing for the highest spin states at 2 GeV and
 2.3 GeV bands. Consequently a prediction was made that the missing states should
 in reality exist and the pattern for the $J=4$ mesons at 2 GeV should be similar
 to the pattern of $J=2$ mesons, while the pattern of $J=5$ states at 2.3 GeV should
 be the same as the pattern of $J=3$ mesons. The chiral  $SU(2)_L \times SU(2)_R$ and $U(1)_A$
 symmetries cannot explain a degeneracy of mesons with different spins. Such a degeneracy
 can be obtained if one assumes a principal quantum number $ \sim n + J$
 on top of $SU(2)_L \times SU(2)_R$ and $U(1)_A$ restorations \cite{GN1}.

The alternative possibility would be to explain the large degeneracy
as due to a principal quantum number $ \sim n + L$, if a
classification of states and assignments of their angular momenta
quantum numbers are done according to standard nonrelativistic
two-body quantum mechanical problem with the LS coupling scheme
\cite{AF2,KZ,SV}. In this case every state is characterized by a set
of three independent conserved angular momenta $^{2S+1}L_J$
\cite{A1,A2,A3,A4,BUGG}. It is easy to see that  there
must not be any parity partners for the highest spin states in
every band within this scenario.
 Such  a degeneracy with the principal quantum number $ \sim n + L$ exists
 in the nonrelativistic Hydrogen atom if one neglects a small spin-orbit force.
 The degeneracy is due to a very specific "accidental" symmetry of the {\it Coulomb} $\sim 1/r$
 interaction in a two-body system. It is hard, however, to imagine that the high
 lying states are driven  by a simple Coulomb part of the one-gluon exchange
 between the constituent quarks. In addition, the Coulomb problem does not
 exhibit any Regge like behavior both for the angular and radial trajectories.

The Nambu-Goto bosonic string type picture implies that the ends of
the string are moving at the speed of light. If one identifies the
ends of the string with the valence quarks   then the valence quarks must be
ultrarelativistic and consequently must have a definite chirality.
Chiral symmetry is not broken. Consequently all states must appear
in chiral multiplets \cite{G2}.

A presence or absence of the chiral partners for the highest spin
states is a key feature that distinguishes both scenarios.
Consequently any reliable experimental information on this issue
would be of fundamental importance. At the moment such states are
not present in the analysis of data \cite{A1,A2,A3,A4,BUGG}. In the
present Brief Report we reanalyze the LEAR data and suggest some
evidence for existence of the missing $\eta_4$ state around 2 GeV.

\section{Analysis of the proton-antiproton annihilation in flight}

The $4^{-+}$ states do not decay into two-pseudoscalar meson
channels or into channels with a neutral pseudoscalar meson and
omega. Therefore these states should be identified from the
reactions with at last three pseudoscalar mesons in final states. Up
to now there are no observations of any $4^{-+}$ states from
analyses of $\pi N$ collision reactions at large energies of
incident pion. The reason can be that these states are produced only
at large energy transferred where statistics is rather low and
partial wave analysis is a rather complicated procedure. The
analysis of the proton-antiproton annihilation in flight into
$\pi^0\pi^0\eta$ channel \cite{A1} observed a $4^{-+}$ isosinglet
state in the region 2.3 GeV  but did not reveal any $4^{-+}$ state
with mass around 2 GeV.

If effective $SU(2)_L \times SU(2)_R$ and $U(1)_A$ trestorations are
correct, then there must be four approximately degenerate mesons
$f_4,a_4,\eta_4,\pi_4$ that are members of the $(1/2,1/2)_a +
(1/2,1/2)_b$ representation \cite{G2,G5}. Let us shortly discuss
properties of the $f_4(2050)$ resonance. The $f_4(2050)$ was
observed very clearly in the $\pi N\to \pi\pi N$ reaction (GAMS
\cite{Alde:1998mc}, BNL\cite{Anisovich:2009zza}), in $\pi N\to
\eta\eta n$ (GAMS \cite{Binon:2004yd}) in $\pi N\to \omega\omega n$
(VES \cite{Amelin:2006wg}), in proton-antiproton annihilation in
flight into two pseudoscalar mesons \cite{Anisovich:2000af}, and in
a set of reactions with three or more mesons in final state (see
\cite{PDG}). The mass of this state is located between 1950-2020 MeV
and the two pion branching is $17\pm 1.5$\%. This resonance
practically does not decay into final $4\pi^0$  state and therefore
about 80 percents of the width should be defined by the decay into
$2\pi\eta$ and charged modes. This resonance contributes about 10\%
to the total cross section of the proton antiproton annihilation
into $\pi^0\pi^0\eta$ final state integrated over mass region
1950-2300 MeV \cite{A1} and decays dominantly into the
$a_2(1320)\pi$ final state.

The $4^{++}$ state can be produced in the proton-antiproton
annihilation either in $^3F_4$ or $^3H_4$ partial waves. The
$4^{-+}$ state can be produced only from $^1G_4$ partial wave and
therefore it should be suppressed by the $\bar p p$ centrifugal
barrier in comparison to the $^3F_4$ amplitude. The kinematical
suppression factor is proportional to the relative momentum of the
initial particles squared calculated in c.m.s of the reaction which
at energy 2 GeV is equal to 0.12 GeV$^2$. The analysis of the
proton-antiproton annihilation in flight showed that the resonance
production vertices are described better with the Blatt-Weiskopf
form factor. In this case the production vertex has a centrifugal
factor:
\be
cf_L=\frac{k^{2L}}{F(L,r^2,k^2)}
\label{factor}
\ee
Here $k$ is the relative momentum of antiproton calculated in c.m.s.
of the reaction, $L$ is the orbital momentum and $r$ is the
effective resonance radius. For $L=3$ and $L=4$ the form factor has
the following form:
\be
F(3,r^2\!,k^2)&\!=\!&\frac{225}{r^6}\!+\!\frac{45k^2}{r^4}\!+\!\frac{6k^4}{r^2}\!+\!k^6
\nn
F(4,r^2\!,k^2)&\!=\!&\frac{11025}{r^8}\!+\!\frac{1575k^2}{r^6}\!+\!\frac{135k^4}{r^4}\!+\!
\frac{10k^6}{r^2}\!+\!k^8
\ee
At 2 GeV the ratio of centrifugal factors $cf_4/cf_3$ is equal to
$\sim 0.06$ for the resonance radius 0.8$fm$ and $\sim 0.13$ for the
radius 1.2$fm$. Due to this centrifugal suppression the $\eta_4$ resonance
around 2 GeV cannot produce any peak in this region,
even if it exists. Note that this suppression applies not only to the
possible $\eta_4$ , but also to all other missing states around 2 GeV,
$\rho_4,\pi_4,\omega_4$, because they are produced in the $L=4$ partial
wave. Similar suppression exists for the missing $J=5$ states in the
2.3 GeV band.

 However the $cf_4$ factor increases much faster with
energy than $cf_3$ and at 2.3 GeV this ratio is equal to 0.25 for the
radius 0.8$fm$ and 0.60 for the radius 1.2$fm$. Thus, if the
production couplings of the $4^{-+}$ and $4^{++}$  states with mass
around 2 GeV are equal
 to each other  as well as branching ratios to the final channel,
the $4^{-+}$ state should contribute between 6-13\% from its
$4^{++}$ partner at energies around 2 GeV and between 25-60\% at
energies around 2.3 GeV.

\section{Fit of the data}

\begin{figure}[pt]
\includegraphics[width=0.52\textwidth,clip=on]
{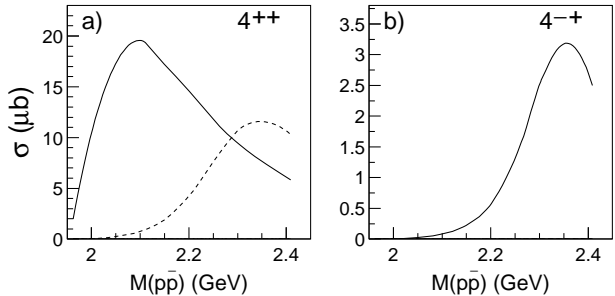}
\caption{\label{sol_main} Contribution of a) the $4^{++}$ states and
b) $4^{-+}$ $\eta_4(2328)$ state to the $\bar p p\to 2\pi^0\eta$ cross section for
the best solution.  In (a) the contribution from $f_4(2050)$ is
given by solid line and the contribution from $f_4(2300)$ as dashed
line.}
\end{figure}

For the  unpolarized proton-antiproton data  the $4^{-+}$ amplitude does
not interfere with either $2^{++}$ or $4^{++}$ amplitudes. However a
$4^{-+}$ state can interfere with $0^{-+}$ and $2^{-+}$ amplitudes
which are dominant contributions to the $\bar p p\to \pi^0\pi^0\eta$
cross section in the mass region around 2 GeV. Therefore one can
hope that even a small contribution of the $4^{-+}$ partial wave can
be identified from such interference. The solution reported in
\cite{A1} and investigated in \cite{b2} found a rather small $4^{-+}$ partial wave which was
described by the Breit-Wigner resonance with mass $2328 \pm 38$ MeV and width
$240 \pm 90$ MeV. The contributions from the $4^{++}$
$f_4(2050)$
and $4^{-+}$  $\eta_4(2328)$ states to
the $\bar p p\to 2\pi^0\eta$ cross section integrated over all decay
modes are shown in Fig.~\ref{sol_main}. In this solution the
$4^{-+}$ partial wave is suppressed by order of magnitude stronger
below 2.2 GeV than what is expected from the centrifugal barrier
factors, which is not very natural.

\begin{figure}
\includegraphics[width=0.52\textwidth,clip=on]
{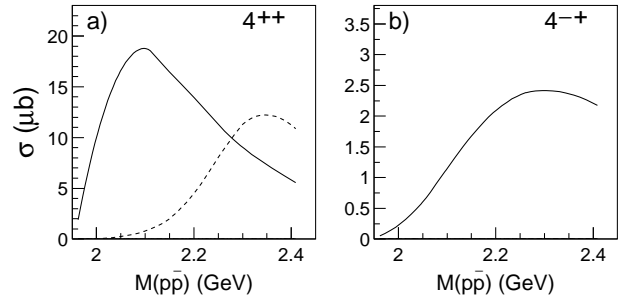}
\caption{\label{sol_low} Contribution of a) the $4^{++}$ states and
b) $4^{-+}$ state to the $\bar p p\to 2\pi^0\eta$ cross section for
the solution with $\eta_4(1950)$.  In (a) the contribution from
$f_4(2050)$ is given by solid line and the contribution from
$f_4(2300)$ as dashed line.}
\end{figure}

Now we want to see what will happen if we substitute the
$\eta_4(2328)$  $4^{-+}$ resonance by a state with  mass 1980 MeV
and allow  decays of this state into $f_2(1275)\eta$,
$a_2(1320)\pi$, $a_0(980)\pi$, $\sigma\eta$ and $f_0(1500)\eta$
channels. The radius for the centrifugal factor was fixed to be 0.8,
1.0, 1.2 and 1.4 fm and width of the resonance was parameterized as
a constant or as a dynamical width defined by the decay into the
$a_0(980)\pi$ channel. The optimization procedure produced an
acceptable likelihood value with $M=1950$ MeV, $\Gamma=380$ MeV,
$r=1.2 fm$ for the parameterization with constant width and $M=1980$
MeV , $\Gamma=360$ MeV, $r=1.2$ fm for the parameterization with
$a_0(980)\pi$ width. The result hardly changed with $r=1.4$ fm. The
contributions from the $4^{++}$ and $4^{-+}$ states to the $\bar p
p\to 2\pi^0\eta$ cross section for the solution with constant width
is shown in Fig.~\ref{sol_low}. It is evident from
Fig.~\ref{sol_low} that in this case there is no unnatural
additional (beyond the centrifugal) suppression of the cross-section
below 2.2 GeV. We also introduced a more complicated
parameterization of the numerator for the $4^{-+}$ state in the form
$a+b\sqrt{s}$. However the $b$ parameter for  such a weak signal only
created a convergency problem and finally was fixed to be zero.

Although this result looks rather promising one should take it with
a caution. First, the total likelihood for this solution was found
to be -89406 which is worse by 135 than that for the best solution,
which is not a significant amount, however. Second, only the lowest
set of data for antiproton beam at 600 MeV was described with a
slightly better likelihood compared to the best solution. Let us
mention that this lowest set has a mass gap of 86 MeV with the
second data set while all other data sets have gaps about 50 MeV.

\begin{figure}
\includegraphics[width=0.32\textwidth,clip=on]
{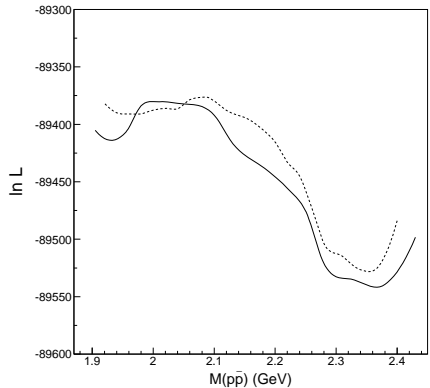} \caption{\label{mass_scan} The mass scan of the
$4^{-+}$ state obtained by changing mass in steps and optimizing of
all other parameters. The solid curve corresponds to the constant
width parameterization and the dashed curve to the resonance width
parameterized as $a_0(980)\pi$ channel.}
\end{figure}

The mass scan of the $4^{-+}$ state for the two width
parameterizations and $r=1.2$ fm is shown in Fig.~\ref{mass_scan}.
It is seen that the distribution of the likelihood value (logarithm
likelihood) has two minima in the mass region investigated. The
minimum at the region 2330 MeV is rather well defined, while the
minimum at 1950-1980 MeV is less pronounced.
 If a state at 1950 MeV is introduced as additional to
$\eta_4(2328)$ the likelihood did not show any improvement due to a
convergency problem. There is no surprise here since this partial
wave provides too small contribution to allow us a complicated
parameterization. However this result does not contradict to the
assumption that both states around 1950 MeV and 2330 MeV are
present.

Let us discuss shortly  properties of the solution with
$\eta_4(1950)$ (see Fig.~\ref{sol_low}). The $4^{-+}$ partial wave
contribution at 2.3 GeV is about 30\% from the $f_4(2050)$ state
which corresponds to  suppression imposed by the centrifugal
barrier with radius 0.8 $fm$. However the situation is a more
complicated here. The main decay mode of $\eta_4(1950)$ was found to
be $a_0(980)\pi$ which is forbidden for $4^{++}$ states. This can
explain a larger width of the $4^{-+}$ state, however the product of
$\bar p p$ and the $a_2(1320)\pi$ couplings (which is the dominant
decay mode for $f_4(2050)$) was optimized to be about 5 times
smaller than that for $f_4(2050)$.

To check whether $4^{-+}$ state at 1950 MeV can be described with
the similar couplings as $f_4(2050)$ we fixed the absolute values
for the couplings into $\bar p p$, $f_2(1275)\eta$ and
$a_2(1320)\pi$ channel to be equal to the lowest orbital momentum
couplings of $f_4(2050)$. After optimization of other parameters we
found the solution which was by 720 worse than the solution with
free couplings. Systematical deviations were seen in description
of angular distributions. Then we decreased the $f_2(1275)\eta$ and
$a_2(1320)\pi$ couplings step by step and obtained a more or less
acceptable solution with a factor 1.5 suppression for the
$f_2(1275)\eta$ channel and 2 for the $a_2(1320)\pi$ channel.

\section{Conclusion}

Although a $4^{-+}$ state at mass 2 GeV was not observed in the
analysis of the proton-antiproton data in flight \cite{A1} there is
a question whether such state could escape identification due to
the centrifugal suppression in the production channel. Indeed, our mass
scan  suggests some indication for existence of the $4^{-+}$ state
with mass about 1950 MeV and width about 380 MeV.

This possible evidence for the missing $\eta_4$ meson around  2 GeV
invites further detailed studies of this missing state.
This state can be confirmed (or disproved) from analysis of new data
on the proton-antiproton annihilation taken from the threshold with
a small step of beam momentum. It would be important to measure not
only neutral final states but also charged modes, in particular
$K^*K$ decay. Due to a suppression in the $\bar p p$
coupling this state should also be searched in different
production reactions, in particular in the $\pi N$
collision at large energies of the incident pion, and in central
collision experiments. Other important question is to initiate a search
of other missing high-spin states around 2 and 2.3 GeV bands, which are
also a subject to the centrifugal suppression in $\bar p p$. Similarly,
a lot of missing states should be found at the 1.7 GeV level.

\medskip
{\bf Acknowledgements}
L.Ya.G. acknowledges support of the Austrian Science
Fund through the grant P21970-N16.

\end{document}